\documentclass[preprint,preprintnumbers,amsmath,amssymb]{revtex4}

\usepackage{graphicx}
\usepackage{dcolumn}
\usepackage{bm}

\begin{document}

\title{To the theory of the Universe evolution}

\author{Boris E. Meierovich}

\affiliation{P.L.Kapitza Institute for Physical Problems.  \\ 2 Kosygina street, Moscow 119334, Russia}
\homepage{http://www.kapitza.ras.ru/people/meierovich/Welcome.html}

\date{\today}

\begin{abstract}
Self-consistent account of the most simple non-gauge vector fields leads to a broad spectrum of regular scenarios of temporal evolution of the Universe completely within the frames of the Einstein's General relativity. The longitudinal non-gauge vector field is ``the missing link in the chain'', displaying the repulsive elasticity and allowing the macroscopic description of the main features of the Universe evolution. The singular Big Bang turns into a regular inflation-like state of maximum compression with the further accelerated expansion at late times. The parametric freedom of the theory allows to forget the troubles of fine tuning. In the most interesting cases the analytical solutions of the Einstein's equations are found.
\end{abstract}

\maketitle

\section{\label{Introduction}Introduction}

Regular scenarios of the Universe evolution, driven by the non-gauge vector fields together with the ordinary matter, are considered within the frames of Einstein's theory of general relativity.
	
From the standpoint of general relativity the matter curves the space-time, giving rise to mutual attraction between the bodies. However, according to modern observations, the Universe is expanding as a whole, despite the gravitational attraction between material objects. The expanding solution of\ the Einstein's equations due to the cosmological constant belongs to De Sitter \cite{De Sitter}. The expanding solutions of\ the Einstein equations without the cosmological constant (Friedman-Robertson-Walker (FRW)\cite{FRW cosmology}) inevitably contained the singularity. The singularity works as a cover for the unknown hidden origin of expansion of the Universe, containing only mutually attracting material objects. For a long time the singularity is considered as a general property of the Universe. The singular point, referred to as ''Big Bang'', is commonly accepted as the ''date of birth'' of the Universe. Discovery of the accelerated expansion of the Universe shows that the source of acceleration continues to exist for a long time after the Big Bang.\ Naturally, the fact of accelerated expansion gave rise to the assumption that the physical vacuum is not just the absence of the ordinary matter. The existence of the so called ''dark energy'' and ''dark matter'', as the unknown source of the Universe expansion, is widely discussed in modern literature \cite{Wikipedia}.
	
Among numerous attempts to guess the riddle of accelerated expansion, from my point of view, the most attractive one is the macroscopic description of the Universe expansion, driven by vector fields. Utilization of vector fields in general relativity shows undoubtable advantages in comparison with scalar fields and with multiplets of scalar fields. The equations appear to be more simple, while their solutions are more general. The solutions have additional parametric freedom, allowing to forget the fine-tuning problem \cite{Meier1}. However, starting from the pioneer paper by Dolgov\cite{Dolgov}, people considered mostly gauge vector fields\cite{Rubakov}-\cite{Koivisto} in applications to the dark sector.
	
Historically in flat space-time the divergence of the vector field was artificially set to zero \cite{Bogolubov-Shirkov} in applications to elementary particles. This restriction allowed to avoid the difficulty of negative contribution to the energy. In the electromagnetic theory it is referred to as Lorentz gauge. The gauge invariance takes place only for massless fields. The Lorentz gauge does not allow to utilize all the advantages of vector fields. In general relativity (in curved space-time) the energy is not a scalar, and its sign is not invariant against the arbitrary coordinate transformations. Considering the vector fields in general relativity, it is worth rejecting the gauge restriction, using instead a more weak condition of regularity. Step by step, people are now getting rid of the Lorentz gauge restrictions \cite{Jimenez1}-\cite{Zuntz}.
	
Vector fields in general relativity form a three-parametric variety \cite{Meier1}. The analysis of the vector fields in the background of the De Sitter metric, including those with a nonzero covariant divergence, is performed in \cite{Meier3}. It came out that the energy-momentum tensor of the most simple zero-mass vector field enters the Einstein equations as an additive to the cosmological constant. Its back reaction affects the Hubble constant -- the rate of expansion. The zero-mass vector field does not vanish in the process of expansion. On the contrary, massive fields vanish with time. Though their amplitude is falling down, the longitudinal massive fields make the expansion accelerated \cite{Meier3}. In other words, the longitudinal massive fields with a nonzero covariant divergence display the repulsive elasticity. It is worth analyzing the possible scenarios of the Universe evolution under the joint action of repulsing vector fields and attracting ordinary matter. Dynamical competition of repulsing and attracting forces results in the variety of regular scenarios of the Universe evolution.
	
The paper is organized as follows. General properties of vector fields in the background of an arbitrary metric are presented in Section{\label{section 2} II}. In Section{\label{section 3} III} we present and analyze the Einstein equations, describing the evolution of the Universe under the action of the most simple non-gauge vector fields. The zero-mass vector field is responsible for either contraction, or expansion, at a constant rate. The massive vector fields grow in the process of contraction and vanish with expansion. They are responsible for the regular transition from contraction to expansion, thus demonstrating the repulsive elasticity and ability to resist compression. In the limits of large and small masses the regular solutions are analyzed analytically. This analysis facilitates considering the regular scenarios of the Universe evolution under the joint action of repulsing vector fields and attracting ordinary matter (Section{\label{section 4} IV}). In most interesting cases the regular solutions are found analytically. The existence of four arbitrary dimensionless parameters allow to forget the fine tuning problem. In the limit of no vector fields the regular analytical solutions reproduce the Friedman-Robertson-Walker \cite{FRW cosmology} fine tuned singular cosmology.
	
The results are discussed in Section{\label{section 5} V}.

\section{\label{section 2} Vector field in general relativity}

In general relativity, the Lagrangian of a vector field $\phi _{I}$ consists of the scalar bilinear combinations of its covariant derivatives and a scalar potential $V(\phi ^{K}\phi _{K})$. A bilinear combination of the covariant derivatives is a 4-index tensor $S_{IKLM}=\phi _{I;K}\phi _{L;M}.$ The most general form of the scalar $S$, formed via contractions of $S_{IKLM}$, is $S=(ag^{IK}g^{LM}+bg^{IL}g^{KM}+cg^{IM}g^{KL})S_{IKLM},$ where $a,b,$ and $c$ are arbitrary constants. The general form of the Lagrangian of a vector field $\phi _{I}$ is
\begin{equation}\label{Lagrangian}
    L=a(\phi _{;M}^{M})^{2}+b\phi _{;M}^{L}\phi _{L}^{;M}+c\phi _{;M}^{L}\phi _{;L}^{M}-V(\phi _{M}\phi ^{M}).
\end{equation}
The classification of vector fields $\phi _{I}$ is most convenient in terms of the symmetric $G_{IK}=\frac{1}{2}(\phi _{I;K}+\phi _{K;I})$ and antisymmetric $F_{IK}=\frac{1}{2}(\phi _{I;K}-\phi _{K;I})$ parts of the covariant derivatives. The Lagrangian (\ref{Lagrangian}) gets the form
\begin{equation}\label{Lagrangian sym and antisym}
    L=a(G_{M}^{M})^{2}+(b+c)G_{M}^{L}G_{L}^{M}+(b-c)F_{M}^{L}F_{\text{ }L}^{M}-V(\phi _{M}\phi ^{M}).
\end{equation}
The bilinear combination of antisymmetric derivatives $F_{M}^{L}F_{\text{ }L}^{M}$ is the same as in electrodynamics. It becomes clear in the common notations $A_{I}=\phi _{I}/2,$ $F_{IK}=A_{I;K}-A_{K;I}.$
	
The terms with symmetric covariant derivatives deserve special attention. In applications to elementary particles in the flat space-time the divergence $\frac{\partial \phi ^{K}}{\partial x^{K}}$ is artificially set to zero \cite{Bogolubov-Shirkov}:
\begin{equation}\label{div fi =0}
    \frac{\partial \phi ^{K}}{\partial x^{K}}=0.
\end{equation}
This restriction allows to avoid the difficulty of negative contribution to the energy. In the electromagnetic theory it is referred to as Lorentz gauge. However, in general relativity (in curved space-time) the energy is not a scalar, and its sign is not invariant against the arbitrary coordinate transformations. Considering the vector fields in general relativity, it is worth getting rid of the restriction (\ref{div fi =0}), using instead a more weak condition of regularity.
	
The covariant field equations
\begin{equation}\label{Covar field eqs}
    a\phi _{;K;I}^{K}+b\phi _{I;K}^{;K}+c\phi _{;I;K}^{K}=-V^{\prime }\phi _{I}
\end{equation}
and the energy-momentum tensor
\begin{equation}\label{T_IK= general}
\begin{array}{l}
T_{IK}=-g_{IK}L+2V^{\prime }\phi _{I}\phi_{K}+2ag_{IK}(\phi_{;M}^{M}\phi^{L})_{;L}  +2(b+c)[(G_{IK}\phi^{L})_{;L}-G_{K}^{L}F_{IL}-G_{I}^{L}F_{KL}] \\
+2(b-c)(2F_{\text{ \ }I}^{L}F_{LK}-F_{\text{ \ }K;L}^{L}\phi_{I}-F_{\text{ \ }I;L}^{L}\phi_{K})
\end{array}
\end{equation}
describe the behavior of vector fields in the background of any arbitrary given metric $g_{IK}$ \cite{Meier2}. Here $V^{\prime }\equiv \frac{dV(\phi _{M}\phi ^{M})}{d(\phi _{M}\phi ^{M})}$. \qquad
	
If the back reaction of the field on the curvature of space-time is essential, then the metric obeys the Einstein equations
\begin{equation}\label{Einstein equations  General}
    R_{IK}-\frac{1}{2}g_{IK}R+\Lambda g_{IK}=\varkappa T_{IK}
\end{equation}
with (\ref{T_IK= general}) added to $T_{IK}.$ Here $\Lambda $ and $\varkappa $ are the cosmological and gravitational constants, respectively. With account of back reaction the field equations (\ref{Covar field eqs}) are not independent. They follow from the Einstein equations (\ref{Einstein equations General}) with $T_{IK}$ (\ref{T_IK= general}) due to the Bianchi identities. The field equations (\ref{Covar field eqs}) are linear with respect to $\phi $ if the vector field is small, and the terms with the second and higher derivatives of the potential $V\left( \phi _{M}\phi ^{M}\right) $ can be omitted.

\section{\label{section3}Vector fields in cosmology}

Today it is generally accepted that among the staff of the Universe only 4.5\% comes from the ordinary matter\cite{NASA}. It is reasonable to start analyzing the role of vector fields in cosmology without the ordinary matter, and include the matter into consideration after the main features of vector fields are clarified.
	
According to observations the Universe expands, and its large scale structure remains homogeneous and isotropic. Consider the space-time with the structure $T^{1}E^{d_{0}}$ and the metric
\begin{equation}\label{Cosmological metric}
    ds^{2}=g_{IK}dx^{I}dx^{K}=(dx^{0})^{2}-e^{2F(x^{0})}\sum_{I=1}^{d_{0}}(dx^{I})^{2}
\end{equation}
depending on only one time-like coordinate $x^{0}=ct$\cite{FOOTNOTE}. Here $d_{0}=3$ is the dimension of space. However, the derivations below are applicable for arbitrary $d_{0}>1.$ The metric tensor $g_{IK}$ is diagonal. The uniform and isotropic expansion is characterized by the single metric function $F(x^{0}),$ and the rate of expansion is $\frac{dF}{dx^{0}}\equiv F^{\prime }$. The Ricci tensor is also diagonal:
\begin{equation}\label{Ricci R_00=}
    R_{00}=-d_{0}(F^{\prime 2}+F^{\prime \prime }),
\end{equation}
\begin{equation}\label{Ricci R_11=}
    R_{II}=e^{2F}(F^{\prime \prime }+d_{0}F^{\prime 2}),\qquad I>0.
\end{equation}

The detailed analysis of vector fields in the cosmological metric (\ref{Cosmological metric})\cite{Meier3} showed that the simplest case \begin{equation}\label{case a}
    a\neq 0,\quad b=c=0
\end{equation}
is the most interesting one from the point of view of the Universe evolution.
	
Depending on the sign of the invariant $\phi ^{I}\phi _{I}\ $the vector $\phi ^{I}$ is either time-like $(\phi ^{I}\phi _{I}>0),$ or space-like $(\phi ^{I}\phi _{I}<0).$ In general relativity, while all coordinates are formally equivalent, one can choose the appropriate coordinate system where $\phi ^{I}=0$ either for $I>0,$ or for $I=0.$ But it can not be done if the coordinate system is already chosen in accordance with some other reasons. In the cosmological metric (\ref{Cosmological metric}) the coordinate $x^{0}$ is already specified, and whatever the sign of the scalar $\phi ^{I}\phi _{I} $ is, we have to consider $\phi _{I}$ having both space and time components.
	
However, all space coordinates in the metric (\ref{Cosmological metric}) are equivalent, and we can choose the coordinate $x^{1}$ along the space direction of the vector field. Then the vector $\phi _{I}$ has only two nonzero components $\phi _{0},$ and $\phi _{1}.$ All other space components of the vector $\phi _{I}$ are zeros:
\begin{equation}\label{fi_I>1 = 0}
    \phi _{I}=0,\text{ \ }I>1.
\end{equation}
	
In the case (\ref{case a}) the vector field equations (\ref{Covar field eqs}) reduce to
\begin{equation}\label{dFi/dx^I=... I=0,1}
    \frac{\partial \Phi }{\partial x^{I}} =-V^{\prime }\phi _{I},\quad I=0,1.
\end{equation}
\begin{equation}\label{Fi=a(...)  case a}
    \Phi \equiv a\phi _{;L}^{L}=a\left( \frac{\partial \phi _{0}}{\partial x^{0}}+d_{0}F^{\prime }\phi _{0}-e^{-2F}\frac{\partial \phi _{1}}{\partial x^{1}}\right) .
\end{equation}
The energy-momentum tensor (\ref{T_IK= general}) reduces to
\begin{equation}\label{T_IK case a}
    T_{IK}=g_{IK}\left( \Phi ^{2}/a+V\right) +2V^{\prime }\left( \phi _{I}\phi _{K}-g_{IK}\phi _{L}\phi ^{L}\right) .
\end{equation}
As usual, the scalar
\begin{equation*}
    V^{\prime }(0)/a=m^{2}
\end{equation*}
in the case (\ref{case a}) can be designated as the square of mass of a vector field.

\subsection{\label{sec31}Massless field}

The equations (\ref{dFi/dx^I=... I=0,1}) for a massless field, $m=0,$ are simply
\begin{equation}\label{field equation case a}
    \frac{\partial \phi _{;L}^{L}}{\partial x^{I}}=0.
\end{equation}
The divergence of the vector field $\phi _{;L}^{L}$ is a constant scalar:
\begin{equation}\label{fi^L_;L = Fi_0/a}
    \phi _{;L}^{L}=\frac{\Phi _{0}}{a},\quad m=0.
\end{equation}

The energy-momentum tensor (\ref{T_IK case a}),
\begin{equation}\label{T_IK    case a m=0}
    T_{IK}=g_{IK}\left( \Phi _{0}^{2}/a+V_{0}\right) ,\quad V^{\prime }=0,
\end{equation}
acts in the Einstein equations (\ref{Einstein equations General}) as a simple addition to the cosmological constant:
\begin{equation}\label{Einstein eqs b=c=V'=0}
    R_{IK}-\frac{1}{2}g_{IK}R+\widetilde{\Lambda }g_{IK}=0,\quad \widetilde{\Lambda }=\Lambda -\varkappa \left( \Phi _{0}^{2}/a+V_{0}\right) .
\end{equation}
Here $V_{0}$ is the value of the constant potential $V\left( \phi _{L}\phi ^{L}\right) $ in the case of massless field $\left( V^{\prime }=0\right) $. The contribution of the zero-mass field to the curvature of space-time remains constant in the process of the Universe evolution.

The metric
\begin{equation}\label{de Sitter metric}
    ds^{2}=(dx^{0})^{2}-e^{\pm \sqrt{-\frac{8\widetilde{\Lambda }}{d_{0}(d_{0}-1)}}(x^{0}-x_{0}^{0})}\sum_{I=1}^{d_{0}}(dx^{I})^{2},\quad d_{0}>1
\end{equation}
is the self-consistent regular solution of the Einstein equations (\ref{Einstein eqs b=c=V'=0}), provided that
\begin{equation}\label{Lambda^<0}
    \widetilde{\Lambda }<0.
\end{equation}
$F(x^{0})$ is a linear function; $x_{0}^{0}$ is a constant of integration. The metric (\ref{de Sitter metric}) is called de Sitter (or anti de Sitter, depending on the sign definition of the Ricci tensor). It describes either expansion (sign $+$), or contraction (sign $-$) of the Universe at a constant rate. In the case of sign $+$ the rate of expansion
\begin{equation}\label{Hubble constant}
    H=\sqrt{-\frac{2\widetilde{\Lambda }}{d_{0}(d_{0}-1)}}
\end{equation}
is called Hubble constant. In our 3-dimensional space
\begin{equation*}
    H=\sqrt{-\frac{1}{3}\widetilde{\Lambda }},\qquad d_{0}=3.
\end{equation*}
	
In general relativity the requirement of regularity (\ref{Lambda^<0}) should replace the artificially imposed restriction (\ref{div fi =0}) which people had been using for a long time in order to avoid the negative energy problem \cite{Bogolubov-Shirkov}.
	
The zero mass vector field determines the constant rate of expansion. Available today properties of the so called dark energy (presently unknown form of matter providing the major contribution to the uniform isotropic expansion of the Universe) can be described macroscopically by the zero-mass vector field with a simple Lagrangian
\begin{equation}\label{Lag case a m=0}
    L=a\left( \phi _{;M}^{M}\right) ^{2}-V_{0}.
\end{equation}
As long as the physical nature of vacuum is not known, the ``geometrical'' origin of the cosmological constant $\Lambda $ and the ``material'' contribution to $\widetilde{\Lambda }$ by the zero-mass vector field can not be separated from one another. The combined action of the massless field and/or the cosmological constant is determined by the single parameter -- Hubble constant (\ref{Hubble constant}).

\subsection{\label{sec32}Massive field}
	
There is a principle difference between the massless and the massive longitudinal vector field. The massive field vanishes with time in the process of expansion, while the massless one does not \cite{Meier3}. If $V^{\prime }\neq 0$ the field equations (\ref{dFi/dx^I=... I=0,1}) allow to express $\phi _{0}$ and $\phi _{1}$ via $\Phi :$
\begin{equation}\label{fi_I=-m^-2...}
    \phi _{I}=-\frac{1}{V^{\prime }}\frac{\partial \Phi }{\partial x^{I}},\quad I=0,1.
\end{equation}
Over the scales much larger than the distances between the galaxies the Universe is uniform and isotropic. Hence, there is no dependence on the space coordinates, and in accordance with (\ref{fi_I=-m^-2...}) $\phi _{1}=0. $ In the scale of the Universe the massive vector field is longitudinal: the only nonzero component $\phi _{0}$ is directed along and depends upon the same time-like coordinate $x^{0}.$
	
The energy-momentum tensor (\ref{T_IK case a}) for the massive field reduces to \cite{Meier3}
\begin{equation}\label{T_mIK}
    T_{IK}=a\left[ g_{IK}\left( \phi _{0}^{\prime }+d_{0}F^{\prime }\phi _{0}\right) ^{2}+\left( 2\delta _{I0}\delta _{K0}-g_{IK}\right) m^{2}\phi _{0}^{2}\right] .
\end{equation}
Here the prime denotes the derivative $d/dx^{0}$ (except that $V^{\prime }=\frac{dV}{d\left( \phi _{L}\phi ^{L}\right) }$). Massive and massless vector fields can be of different physical nature. In the Einstein equations
\begin{equation}\label{Eq  I=0 a not eq 0}
    \frac{1}{2}d_{0}\left( d_{0}-1\right) F^{\prime 2}+\widetilde{\Lambda } =\varkappa a\left[ \left( \phi _{0}^{\prime }+d_{0}F^{\prime }\phi _{0}\right) ^{2}+m^{2}\phi _{0}^{2}\right] ,
\end{equation}
\begin{equation}\label{Eq I>0 a not eq 0}
    \left( d_{0}-1\right) F^{\prime \prime }+\frac{1}{2}d_{0}\left( d_{0}-1\right) F^{\prime 2}+\widetilde{\Lambda } =\varkappa a\left[ \left( \phi _{0}^{\prime }+d_{0}F^{\prime }\phi _{0}\right) ^{2}-m^{2}\phi _{0}^{2}\right]
\end{equation}
the massless field is taken into account via the constant $\widetilde{\Lambda },$ and the massive one is described by the function $\phi _{0}\left( x^{0}\right) $ \cite{Meier3}. Their applicability is restricted by the condition that the second and higher derivatives of the potential $V\left( \phi _{L}\phi ^{L}\right) $ can be ignored. Regular solutions of these equations are possible if $\widetilde{\Lambda }$ and $a$ are of the same sign. It follows from (\ref{Eq I=0 a not eq 0}), provided that $F^{\prime }$ can change the sign.
	
The vector field equations (\ref{Covar field eqs}) reduce to the only one equation
\begin{equation}\label{Vec massive Field eq}
    \left( \phi _{0}^{\prime }+d_{0}F^{\prime }\phi _{0}\right) ^{\prime }+m^{2}\phi _{0}=0,
\end{equation}
which is the consequence of the Einstein equations (\ref{Eq I=0 a not eq 0}-\ref{Eq I>0 a not eq 0}) due to the Bianchi identities. Extracting
(\ref{Eq  I=0 a not eq 0}) \ from (\ref{Eq I>0 a not eq 0}), we have
\begin{equation}\label{F''=  a not equal 0}
    F^{\prime \prime }=-\frac{2a}{d_{0}-1}\varkappa m^{2}\phi _{0}^{2}.
\end{equation}
The fourth order set of equations (\ref{Vec massive Field eq}-\ref{F''= a not equal 0}) with the boundary condition
\begin{equation}\label{boun cond at F'=0}
    \phi _{0}^{\prime 2}+m^{2}\phi _{0}^{2}=\frac{\widetilde{\Lambda }}{\varkappa a}\quad \text{at\quad }F^{\prime }=0,
\end{equation}
following from (\ref{Eq I=0 a not eq 0}), has the same solutions as the third order set (\ref{Eq I=0 a not eq 0}-\ref{Eq I>0 a not eq 0}). In the regular expanding solutions at $x^{0}\rightarrow \infty $ , in accordance with (\ref{Lambda^<0}), $\widetilde{\Lambda }<0.$ Hence, $a$ is also negative, and $F^{\prime \prime }$ (\ref{F''= a not equal 0}) is positive: the massive longitudinal vector field makes the expansion of the Universe accelerated. Without ordinary matter the second derivative (\ref{F''= a not equal 0}) does not change sign. $x^{0}$ is a cyclic coordinate, and it is convenient to set $F^{\prime }=0$ at the origin $x^{0}=0.$ In view of the $x^{0}\rightarrow -x^{0}$ invariance of the equations, it is clear, that the rate of expansion $F^{\prime }\left( x^{0}\right) $ is a monotonically growing function between its limiting values $F^{\prime }\left( -\infty \right) =-H$ in the past and $F^{\prime }\left( +\infty \right) =H$ \ in future. The scale factor $R=e^{F}$ decreases with time while $F^{\prime }<0,$ reaches its minimum, and grows when $F^{\prime }$ becomes positive.
	
One of the two constants $\phi _{0}$ and $\phi _{0}^{\prime }$ at $x^{0}=0$ remains arbitrary within the boundary condition (\ref{boun cond at F'=0}). In the case $\phi _{0}^{\prime }\left( 0\right) =0$ the field $\phi _{0}\left( x^{0}\right) $ is a symmetric function, and in the case $\phi _{0}\left( 0\right) =0$ it is an antisymmetric one. In these both cases $F^{\prime }\left( x^{0}\right) $ is antisymmetric. If both constants $\phi _{0}$ and $\phi _{0}^{\prime }$ at $x^{0}=0$\ are not zeroes, the regular solutions still exist, but there is no symmetry with respect to $x^{0}\rightarrow -x^{0}.$
	
It is convenient to perform the further analytical and numerical analysis in dimensionless variables $\phi $ and $z$:
\begin{equation}\label{dimensionless variables}
    \phi =\sqrt{-\frac{2a\varkappa }{d_{0}\left( d_{0}-1\right) }}\phi _{0},\quad z=Hx^{0}.
\end{equation}
The equations (\ref{Vec massive Field eq},\ref{F''= a not equal 0})
\begin{equation}\label{Eq for fi}
    \left( \phi _{,z}+d_{0}F_{,z}\phi \right) _{,z}+\mu ^{2}\phi  =0,
\end{equation}
\begin{equation}\label{Eq for F}
    F_{,z,z} =d_{0}\mu ^{2}\phi ^{2}
\end{equation}
contain only one dimensionless parameter
\begin{equation}\label{mu=m/H}
    \quad \mu =\frac{m}{H}.
\end{equation}
(In dimensional units $\mu =\frac{mc^{2}}{\hbar H}$). Subscript $_{,z}$ denotes $d/dz.$ The boundary relations (\ref{boun cond at F'=0}) in terms of $\phi \left( 0\right) ,$ $\phi _{,z}\left( 0\right) ,$ and $F_{,z}\left( 0\right) $ are
\begin{equation}\label{boundary relations}
    \phi _{,z}^{2}\left( 0\right) +\mu ^{2}\phi ^{2}\left( 0\right) =1,\quad F_{,z}\left( 0\right) =0.\quad
\end{equation}
$F\left( z\right) $ enters the equations (\ref{Eq for fi},\ref{Eq for F}) only via the derivatives, but not directly. For this reason $F_{0}=F\left( 0\right) $ remains arbitrary. The constant $F_{0}$ only determines the scale of the space coordinates, and it does not change the structure of the metric and vector field.

\subsubsection{\label{sec321}Asymptotic behavior at large $x^{0}$}

At late times the back reaction of the massive field on the metric becomes negligible. In accordance with (\ref{Eq for F}) $F_{,z,z}\rightarrow 0,$ and $F_{,z}\rightarrow 1$ at $x^{0}\rightarrow \infty .$ The late temporal evolution of the massive field is described by (\ref{Eq for fi}) with $F_{,z}=1.$ Its solution
\begin{equation}\label{psi(z) at z to inf}
    \phi \left( z\right) =C_{+}e^{\lambda _{+}z}+C_{-}e^{\lambda _{-}z},\qquad \lambda _{\pm }=-\frac{d_{0}}{2}\pm \sqrt{\left( \frac{d_{0}}{2}\right) ^{2}-\mu ^{2}},\qquad z\rightarrow \infty
\end{equation}
is a linear combination of two functions, vanishing at $z\rightarrow \infty $. The functions are monotonic, if $\mu <$ $\frac{d_{0}}{2},$ or oscillating with a decreasing magnitude, if $\mu >$ $\frac{d_{0}}{2}.$ If $\mu $ is small, the field decreases very slowly: \begin{equation}\label{fi(z) slow at z to inf}
    \phi \left( z\right) =C_{+}\exp \left( -\frac{2\mu ^{2}}{d_{0}^{2}}z\right) ,\qquad z\rightarrow \infty ,\qquad \mu \ll d_{0}.
\end{equation}
	
For large and small $\mu $ the equations (\ref{Eq for fi}-\ref{Eq for F}) can be solved analytically.

\subsubsection{\label{sec322}Large $\mu $}

In the case of large $\mu ,$
\begin{equation}\label{mu >> 1}
    \mu \gg 1,
\end{equation}
the field $\phi \left( z\right) $ is a rapidly oscillating function as compared with $F\left( z\right) .$ We search the solution of the field equation (\ref{Eq for fi}) in the form
\begin{equation*}
    \phi \left( z\right) =\frac{1}{\mu }\psi \left( z\right) \cos \left( \mu z+\varphi \right) .
\end{equation*}
In accordance with (\ref{boundary relations}) $ \psi \left( 0\right) =1,$ and the phase $\varphi \ $ depends on the relation between the boundary values $\phi \left( 0\right) $ and $\phi _{,z}\left( 0\right) .$ The main terms $\sim \mu $ in (\ref{Eq for fi}) disappear, the next order terms $\sim 1$ are:
\begin{equation*}
    2\psi _{,z}+d_{0}F_{,z}\psi =0.
\end{equation*}
The remaining terms $\sim 1/\mu \ll 1$ can be neglected. The solution is
\begin{equation*}
    \phi \left( z\right) =\frac{1}{\mu }e^{-d_{0}\left( F-F_{0}\right) /2}\cos \left( \mu z+\varphi \right) ,\quad \mu \gg 1.
\end{equation*}
Here $F_{0}=F\left( 0\right) .$

Averaging (\ref{Eq for F}) over the $z$-interval much larger than the period of vibrations $\mu ^{-1},$ we get the following equation for $\overline{F}\left( z\right) :$
\begin{equation*}
    \overline{F}_{,z,z}=\frac{d_{0}}{2}e^{-d_{0}\left( \overline{F}-F_{0}\right) },\quad \mu \gg 1.
\end{equation*}
The solution is
\begin{equation}\label{overline(F)_(z) =}
    \overline{F}\left( z\right) = F_{0}+\frac{2}{d_{0}}\ln \cosh \left( \frac{d_{0}}{2}z\right) ,\quad \overline{F}_{,z}\left( z\right) =\tanh \left( \frac{d_{0}}{2}z\right) ,
\end{equation}
\begin{equation}\label{fi(z) = mu>>1}
    \phi \left( z\right) = \frac{1}{\mu }\frac{\cos \left( \mu z+\varphi \right) }{\cosh \left( \frac{d_{0}}{2}z\right) },\quad \mu \gg 1,
\end{equation}
	
\begin{figure} \centering
\vspace{0cm}
    \hspace{0cm}
  \includegraphics[width=10 cm]{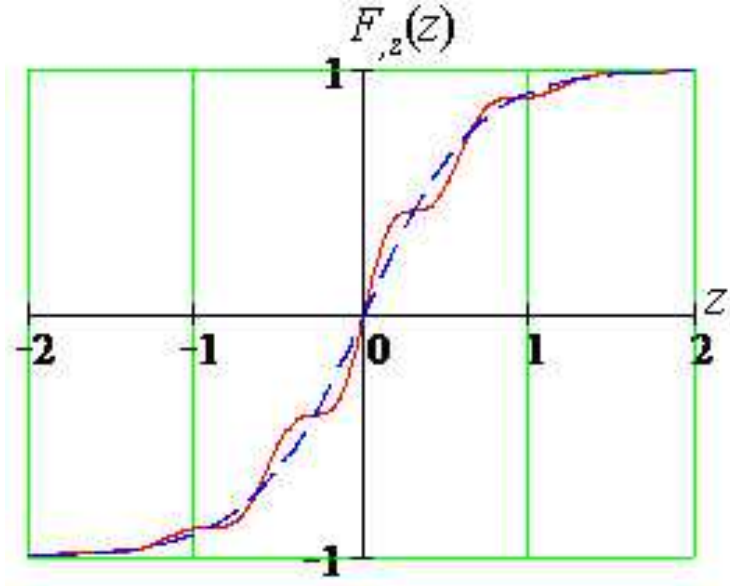}
  \caption{Red curve is the numerical solution $F_{,z}\left( z\right)$ for $\mu =5,$ and $\overline{F}_{,z}\left( z\right) =\tanh \left( \frac{d_{0}}{2}z\right) -$ blue dashed line; $d_{0}=3$.}\label{Fig1}
\end{figure}

Oscillations of $\phi \left( z\right) $\ at large $\mu $ initiate weak vibrations of the function $F\left( z\right) $ around the averaged value $\overline{F}\left( x\right) ,$ see Figure \ref{Fig1}. Red curve is the numerical solution for $\mu =5,$ $\phi _{,z}\left( 0\right) =0.$ Blue dashed line -- analytical solution $\overline{F}_{,z}\left( z\right) $ (\ref{overline(F)_(z) =}).
\begin{figure} \centering
\vspace{0cm}
    \hspace{0 cm}
  \includegraphics[width=16 cm]{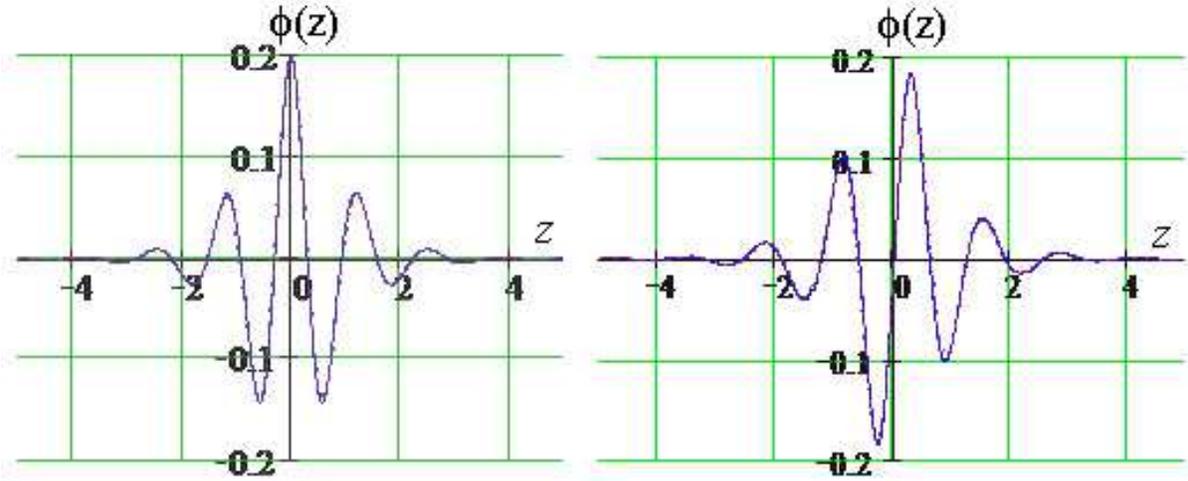}
  \caption{Symmetric (left) and antisymmetric (right) $\phi \left( z\right)$,  found numerically for $\mu =5$ and $d_{0}=3 $, coincide with found analytically (\ref{fi(z) = mu>>1}).}\label{Fig2}
\end{figure}
In Figures 2-left and 2-right $\phi \left( z\right), $ found numerically for $\mu =5\ $, practically coincide with found analytically (\ref{fi(z) = mu>>1}). The boundary condition $\phi _{,z}\left( 0\right) =0$ for the symmetric solution in Figure 2-left corresponds to $\varphi =0$ in (\ref{fi(z) = mu>>1}). The antisymmetric solution in Figure 2-right (the boundary condition $\phi \left( 0\right) =0$) coincides with (\ref{fi(z) = mu>>1}) at $\varphi =-\pi /2.$

\subsubsection{\label{sec323}Case $\mu $ is very small}

If in the case $\mu \ll 1$ we neglect the term $\mu ^{2}\phi $ in the vector field equation (\ref{Eq for fi}), then the vector field $\phi \left( z\right) $ is expressed via $F\left( z\right) $ as follows:
\begin{equation}\label{fi(z) small mu}
    \phi \left( z\right) =\left( \phi _{,z}\left( 0\right) \int_{0}^{z}e^{d_{0}\left[ F\left( z_{1}\right) -F_{0}\right] }dz_{1}+\phi \left( 0\right) \right) e^{-d_{0}\left[ F\left( z\right) -F_{0}\right] },\quad \mu \ll 1.
\end{equation}
The constants $\phi \left( 0\right) $ and $\phi _{,z}\left( 0\right) $ obey the boundary relation (\ref{boundary relations}).

\quad  \\  \quad   \textsl{Symmetric and antisymmetric configurations}
	
In the symmetric case $\phi _{,z}\left( 0\right) =0$ (\ref{fi(z) small mu}) reduces to
\begin{equation}\label{fi(z) symm mu<<1}
    \phi \left( z\right) =\mu ^{-1} e^{-d_{0}\left[ F\left( z\right) -F_{0}\right] },\quad \mu \ll 1.
\end{equation}
$\phi \left( 0\right) =\mu ^{-1}$ in accordance with the boundary relation (\ref{boundary relations}). Neglecting the term $\mu ^{2}\phi $ in the vector field equation (\ref{Eq for fi}) results in $C_{-}=0$ in the asymptotic behavior (\ref{psi(z) at z to inf}) of $\phi \left( z\right) $ at $z\rightarrow \infty .$ On the contrary, omitting\ $\mu ^{2}\phi $ in (\ref{Eq for fi}) in the antisymmetric case $\phi \left( 0\right) =0$ corresponds to $C_{+}=0.$ The integral in (\ref{fi(z) small mu}) converges at $z\rightarrow \infty ,$ and both expressions, (\ref{fi(z) small mu}) and
(\ref{psi(z) at z to inf}), give the same result: $\phi \left( z\right) \rightarrow const$ at $\mu \rightarrow 0$ and $z\rightarrow \infty .$\ Antisymmetric $\phi \left( z\right) $, acting as a massless field\ in the limit $\mu \rightarrow 0,$\ renormalizes the Hubble constant. Therefore it is reasonable to consider the contribution of the antisymmetric $\phi \left( z\right) $ at $\mu \rightarrow 0$\ as already included into $\widetilde{\Lambda }$.
	
In the symmetric case the metric is determined by the equation
\begin{equation*}
    F_{,z,z}=d_{0}e^{-2d_{0}\left[ F\left( z\right) -F_{0}\right] }
\end{equation*}
and boundary conditions
\begin{equation*}
    F_{,z}\left( 0\right) =0,\quad F\left( 0\right) =F_{0}.
\end{equation*}
The analytical solution for the symmetric configuration
\begin{equation}\label{Anal mu<<1}
   F\left( z\right) =F_{0}+\frac{1}{d_{0}}\ln \cosh \left( d_{0}z\right) , \\
   F_{,z}\left( z\right) =\tanh \left( d_{0}z\right) ,  \\
   \phi \left( z\right) =\frac{1}{\mu \cosh \left( d_{0}z\right) },\quad \mu \ll 1,
\end{equation}
describes the transition from the compression to the expansion, see Figure 3.
\begin{figure} \centering
\vspace{0cm}
    \hspace{0 cm}
  \includegraphics[width=16 cm]{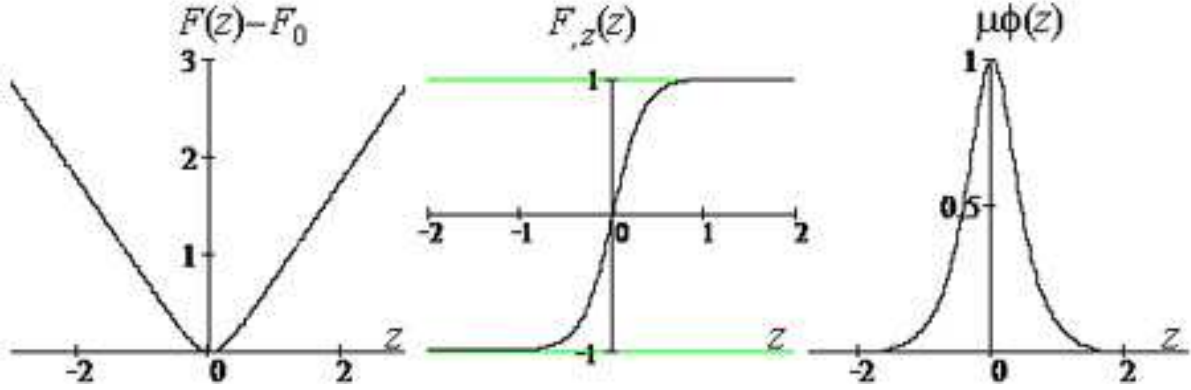}
  \caption{Symmetric solution $\ F\left( z\right) -F_{0},$ $\ F_{,z}\left( z\right),$  $\mu \phi \left( z\right)$ found analytically (\ref{Anal mu<<1}) for  $\mu \ll 1.$ $d_{0}=3$
  }\label{Fig3}
\end{figure}	
	
With no ordinary matter the time interval of transition is of the order of Hubble time $\sim 1/d_{0}H.$ At $\mu \ll 1$\ it does not depend on the mass $m$ of the massive field. The scale factor is
\begin{equation*}
    R\left( z\right) =e^{F\left( z\right) }=e^{F_{0}}\left[ \cosh \left( d_{0}z\right) \right] ^{1/d_{0}}.
\end{equation*}
Without the ordinary matter the acceleration $F_{,z,z}$ is positive:
\begin{equation*}
    F_{,z,z}\left( z\right) =\frac{d_{0}}{\cosh ^{2}\left( d_{0}z\right) }>0,\quad \mu \ll 1.
\end{equation*}
	
Like an elastic spring, the symmetric longitudinal vector field enables the transition from compression to expansion. The kinetic energy of contraction completely converts at $x^{0}=0$ into potential energy of the compressed vector field, and at $x^{0}>0$ the energy is being released back in the form of the kinetic energy of expansion.
	
An example of the antisymmetric solution (with the boundary condition $\phi \left( 0\right) =0$ and $\mu =0.25$), found numerically for $d_{0}=3$, is presented in Figure 4.
\begin{figure} \centering
\vspace{0cm}
    \hspace{0 cm}
  \includegraphics[width=15 cm]{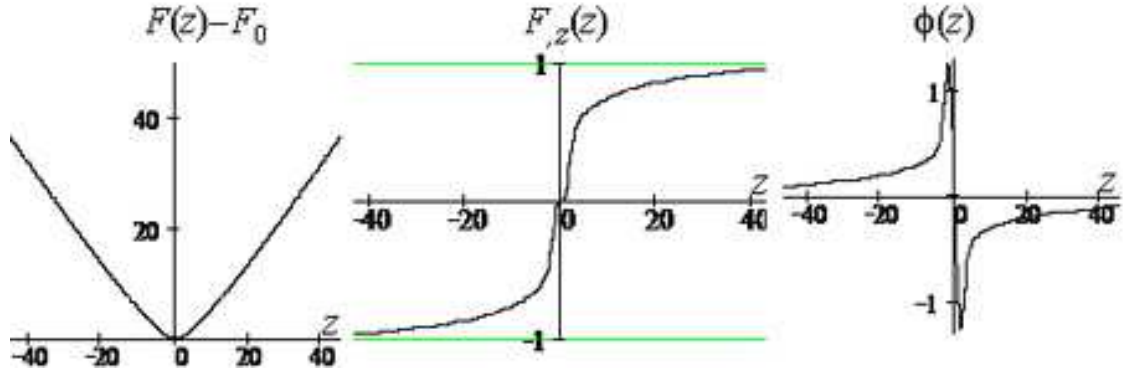}
  \caption{Antisymmetric solution $\ F\left( z\right) -F_{0},$ $\ F_{,z}\left( z\right),$  $\phi \left( z\right),$ found numerically for $\mu=0.25,$  $d_{0}=3$}\label{Fig4}
\end{figure}
	
Numerical analysis confirms, that at $\mu \ll 1$ the antisymmetric $\phi \left( z\right) $ decreases very slowly  as
 at $z\rightarrow \infty$, see (\ref{fi(z) slow at z to inf}) and Figure 4 (right). The symmetric field (Figure 3 right) vanishes quickly, $\sim \exp \left( -d_{0}z\right) $.

\section{\label{section4}Evolution determined by vector fields and ordinary matter}
	
The above analysis of general properties of the equations (\ref{Eq for fi},\ref{Eq for F})\ facilitates clarifying the solutions of the Einstein equations with the ordinary matter taken into account.
	
\subsection{\label{sec41}Dust matter approximation}
	
Applying the general relativity to the Universe as a whole it is natural to consider the ordinary matter (stars, galaxies, ...) as separated noninteracting subsystems located far from one another. Averaged over the distances larger than the distance between the objects, the ordinary matter can be considered macroscopically as a uniformly distributed dust. As far as the ordinary matter does not violate the homogeneity and isotropy of the large scale structure of the space, all the components of its energy-momentum tensor, except $T_{00}=\rho ,$ are zeros:
\begin{equation}\label{T_IK =ro.}
    T_{IK}=\rho \delta _{I0}\delta _{K0}.
\end{equation}
In the process of expansion the averaged energy density of matter $\rho $ depends only on $x^{0}$.
	
$T_{IK}$ is a symmetric tensor. Due to the Bianchi identities its covariant divergence is zero,
\begin{equation}\label{T^K_I;K=}
    T_{I;K}^{K}=\frac{1}{\sqrt{g}}\frac{\partial }{\partial x^{K}}\left( \sqrt{g}T_{I}^{K}\right) -\frac{1}{2}\frac{\partial g_{KL}}{\partial x^{I}}T^{KL}=0.
\end{equation}
In the cosmological metric (\ref{Cosmological metric})\ the covariant divergence (\ref{T^K_I;K=}) of the tensor (\ref{T_IK =ro.})\ is \begin{equation*}
    T_{I;K}^{K}=e^{-d_{0}F}\frac{\partial \left( e^{d_{0}F}\rho \right) }{\partial x^{0}}\delta _{I0}.
\end{equation*}
Thus, $\rho e^{d_{0}F}=const.$ If we accept that $\rho \left( x^{0\ast }\right) =\rho _{0}\text{ }$is the averaged density of the ordinary matter now, then
\begin{equation*}
    \rho \left( x^{0}\right) =\rho _{0}e^{-d_{0}F\left( x^{0}\right) },
\end{equation*}
and the present moment $x^{0\ast }$ is defined by
\begin{equation}\label{F(x^0*)=0}
    F\left( x^{0\ast }\right) =0.
\end{equation}
	
\subsection{\label{sec42}Einstein equations}
	
With the dust matter taken into account, the Einstein equations (\ref{Eq I=0 a not eq 0},\ref{Eq I>0 a not eq 0})\ are:
\begin{equation*}
    \frac{1}{2}d_{0}\left( d_{0}-1\right) F^{\prime 2}+\widetilde{\Lambda } =\varkappa a\left[ \left( \frac{d\phi _{0}}{dx^{0}}+d_{0}F^{\prime }\phi _{0}\right) ^{2}+m^{2}\phi _{0}^{2}\right] +\varkappa \rho _{0}e^{-d_{0}F},
\end{equation*}
\begin{equation*}
    \left( d_{0}-1\right) F^{\prime \prime }+\frac{1}{2}d_{0}\left( d_{0}-1\right) F^{\prime 2}+\widetilde{\Lambda } =\varkappa a\left[ \left( \frac{d\phi _{0}}{dx^{0}}+d_{0}F^{\prime }\phi _{0}\right) ^{2}-m^{2}\phi _{0}^{2}\right] .
\end{equation*}
In dimensionless variables (\ref{dimensionless variables})
\begin{equation}\label{Eq I=0}
    F_{,z}^{2}-1 = -\left( \phi _{,z}+d_{0}F_{,z}\phi \right) ^{2}-\mu ^{2}\phi ^{2}+\Omega e^{-d_{0}F},
\end{equation}
\begin{equation}\label{Eq I>0}
    \frac{2}{d_{0}}F_{,z,z}+F_{,z}^{2}-1 = -\left( \phi _{,z}+d_{0}F_{,z}\phi \right) ^{2}+\mu ^{2}\phi ^{2}.
\end{equation}
The parameter $\Omega ,$
\begin{equation}\label{Omega}
    \Omega =-\frac{\varkappa \rho _{0}}{\widetilde{\Lambda }}=\frac{2\varkappa \rho _{0}}{d_{0}\left( d_{0}-1\right) H^{2}},
\end{equation}
denotes the ratio of the energy density of the ordinary matter to the density of the kinetic energy of expansion. The vector field equation
(\ref{Eq for fi}) remains the same, and instead of the equation (\ref{Eq for F}) we now have
\begin{equation}\label{F_,z,z = ...}
    F_{,z,z}=d_{0}\mu ^{2}\phi ^{2}-\frac{d_{0}\Omega }{2}e^{-d_{0}F}.
\end{equation}
This equation resembles the Newton's law: acceleration $F_{,z,z}$ is proportional to the ``repulsing force'' $d_{0}\mu ^{2}\phi ^{2}$ minus the ``attracting force'' $\frac{d_{0}\Omega }{2}e^{-d_{0}F}.$ Altogether the regular solutions of the set (\ref{Eq for fi},\ref{F_,z,z = ...}) with the boundary conditions
\begin{equation}\label{bound cond with ord matter}
    \phi _{,z}^{2}\left( 0\right) +\mu ^{2}\phi ^{2}\left( 0\right) =1+\Omega e^{-d_{0}F_{0}},\quad F_{,z}\left( 0\right) =0,\quad F\left( 0\right) =F_{0},\quad \widetilde{\Lambda }<0,
\end{equation}
contain five dimensionless parameters: $\mu ,\Omega ,F_{0},\phi \left( 0\right) ,$ and $\phi _{,z}\left( 0\right) .$ In view of the bounding relation (\ref{bound cond with ord matter}), four of them remain independent.

\subsection{\label{sec43}Regular cosmological solutions}
	
The equations (\ref{Eq for fi},\ref{F_,z,z = ...}) with the boundary conditions (\ref{bound cond with ord matter}) are easily integrated numerically. The regular solutions are free from fine tuning. Moreover, the existing parametric freedom results in a great variety of possible configurations. Among the four independent parameters, $\Omega $ (\ref{Omega}) can be estimated better than others. Substituting $d_{0}=3,$ $\varkappa =6.67\times 10^{-8}$ cm$^{3}/$g $\sec ^{2},$ $H=2\times 10^{-18}$ sec$^{-1},$ $\rho _{0}=2\times 10^{-31}$ g/cm$^{3}$ (universal density of luminous matter \cite{Gravitation}) we have $\Omega \simeq 10^{-3}.$ Numerical analysis shows, that if both $\phi \left( 0\right) ,$ and $\phi _{,z}\left( 0\right) $ are not zeros at a turning point $F_{,z}=0,$ then there can be other even more sharp turning points with $\phi _{,z}$ more close to zero. With fixed values of two parameters, $\left( \Omega =10^{-3},\text{ }\phi _{,z}\left( 0\right) =0\right) ,$ the role of $\mu $ at fixed $F_{0}=-3$ in the process of evolution is demonstrated in Figure 5, and the role of $F_{0}$ at fixed $\mu =3$ is shown in Figure 6.
 
\begin{figure}
  \includegraphics[width=7 cm]{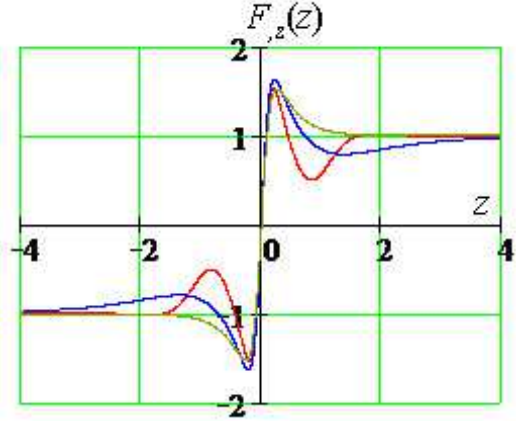}\\
  \caption{The rate of expansion $F_{,z}\left( z\right) .$ $\mu =0,1,2$ -- brown, blue, red curves. $d_{0}=3,$ $\Omega =0.001,$ $F_{0}=-3.$}\label{Fig5}
\end{figure}
The rate of evolution $F_{,z}\left( z\right) $ for $\mu =0,1,$ and $2$ is presented in Figure 5 by brown, blue, and red  curves, respectively. In all symmetric configurations the global transition from contraction to expansion takes place at $z=0.$ At fixed $F_{0}=-3$ after the transition the rate of expansion reaches its first maximum $F_{,z}\left( z_{\max }\right) >1$ at some $z_{\max }>0.$ At $z=z_{\max }$ the acceleration turns into deceleration. If $\mu =0$ (brown curve in Figure 5) $F_{,z}\left( z>z_{\max }\right) $ decreases monotonically to $F_{,z}=1$ as $z\rightarrow \infty .$ If \ $\mu >0$ a next point of extremum $z_{\min }>z_{\max }$ appears, where deceleration turns back into acceleration. The bigger is $\mu ,$ the deeper\ is the minimum of $F_{,z}$ at $z=z_{\min },$ -- compare the blue ($\mu =1$) and red ($\mu =2$) curves in Figure 5. The number of subsequent maxima and minima grows with increasing $\mu .$
	
Parameter $F_{0}$ determines the maximal speed of the outburst $F_{,z}\left( z_{\max }\right) $ in the contraction-to-expansion transition$.$ The rate of expansion $F_{,z}\left( z\right) $ for $F_{0}=-3,-4,$ and $-5$ at fixed $\mu =3$ is presented in Figure 6. The maximum $F_{,z}\left( z_{\max }\right) $ grows exponentially with the increasing negative value of $F_{0}.$ It resembles inflation, except that there is no singularity. Some authors call the regular contraction-to-expansion transition ``nonsingular bounce''\cite{Creminelli}, \cite{Lin},\cite{Steinhardt}. In the literature there are attempts to find a self-consistent model in order to explain from a unified viewpoint the inflation in the early Universe and the late-time accelerated expansion \cite{Balakin}. However, one should keep in mind that the dust matter approximation is applicable until the galaxies are noninteracting systems, located at far distances from one another.
\begin{figure}
  \includegraphics[width=6 cm]{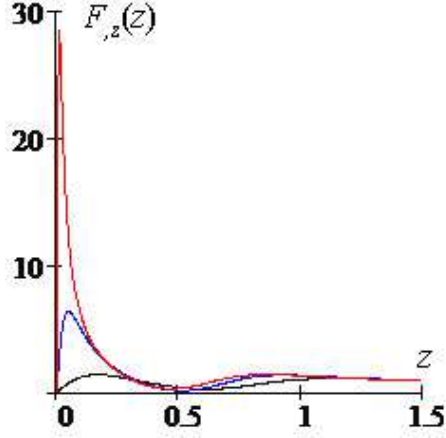}\\
  \caption{The rate of expansion $F_{,z}(z).$ $\mu =3.$ $F_{0}=-3,-4,-5$ -- brown, blue, red curves. $d_{0}=3,$ $\Omega =0.001$}\label{Fig6}
\end{figure}
	
In the case of small $\mu \ll 1$ the transition from contraction to expansion can be described analytically. Substituting (\ref{fi(z) symm mu<<1}) into (\ref{F_,z,z = ...}), we exclude $\phi $ and come to the single equation for $F:$
\begin{equation}\label{Eq for F small mu}
    F_{,z,z}=d_{0}e^{-2d_{0}\left( F-F_{0}\right) }-\frac{d_{0}\Omega }{2}e^{-d_{0}F}.
\end{equation}
Its solution with the boundary conditions (\ref{bound cond with ord matter}) is
\begin{equation}\label{F(z)=  mu<<1}
    F\left( z\right) =F_{0}+\frac{1}{d_{0}}\ln \left[ \left( 1+\frac{1}{2}\Omega e^{-d_{0}F_{0}}\right) \cosh \left( d_{0}z\right) -\frac{1}{2}\Omega e^{-d_{0}F_{0}}\right] ,\quad \mu \ll 1.
\end{equation}
For the rate of evolution $F_{,z}\left( z\right) $ and for the scale factor $R\left( z\right) =e^{F\left( z\right) }$ we get
\begin{equation}\label{F_,z(z)=   mu << 1}
    F_{,z}\left( z\right) =\frac{\sinh \left( d_{0}z\right) }{\cosh \left( d_{0}z\right) -\left( 1+\frac{2}{\Omega }e^{d_{0}F_{0}}\right) ^{-1}},
\end{equation}
\begin{equation}\label{R(z)=   mu << 1}
    R\left( z\right) =\left[ \left( e^{d_{0}F_{0}}+\frac{1}{2}\Omega \right) \cosh \left( d_{0}z\right) -\frac{1}{2}\Omega \right] ^{\frac{1}{d_{0}}}.
\end{equation}
Analytical solutions (\ref{F(z)= mu<<1}-\ref{R(z)= mu << 1}), derived for $\mu \ll 1,$ are as well applicable in the vicinity of the transition for $\mu \sim 1$ if $\left| F_{0}\right| \gg 1$. See Figure 7, where the variation of $F_{,z}\left( z\right) $ in the vicinity of the turning point, found numerically for $\mu =10$ and $F_{0}=-10,$ coincides with (\ref{F_,z(z)= mu << 1}).
\begin{figure}
  \includegraphics[width=6.6711 cm]{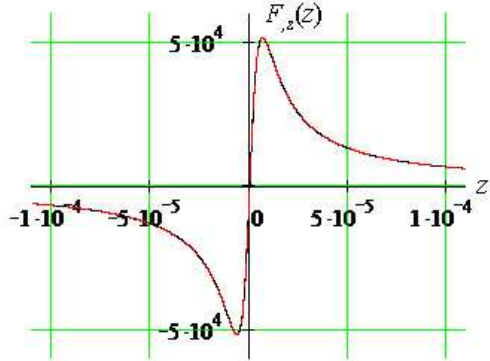}\\
  \caption{Variation of $F_{,z}\left( z\right) $ in the vicinity of the turning point, found numerically for $\mu =10$ and $F_{0}=-10,$ coincides with (\ref{F_,z(z)= mu << 1})}\label{Fig7}
\end{figure}		
It is because for very large negative $F_{0}$ the width of the contraction-to-expansion transition $\Delta z$ is very narrow:
\begin{equation*}
    \Delta z\sim \frac{2}{d_{0}\sqrt{\Omega }}e^{d_{0}F_{0}/2},\quad F_{0}<0,\quad \left| F_{0}\right| \gg 1.
\end{equation*}

According to the recent analysis of the Hubble space telescope data, the expansion of the Universe switched from deceleration to acceleration at about the half of the age of the Universe\cite{Suzuki}. In the analytical solution (\ref{F(z)= mu<<1}) $F_{,z,z}\left( z\right) $ is negative at $z>z_{\max }:$ if $\mu =0,$ the expansion goes with deceleration. The switching from deceleration to acceleration (appearance of the minimum of $F_{,z}\left( z\right) $ at $z=z_{\min },$ see Figures 5,6) is a hint that the symmetric longitudinal vector field $\phi $ is massive, $\mu >0.$ At $\mu >\frac{d_{0}}{2}$ the field $\phi \left( z\right) $ becomes an oscillating function, and the number of minimums of $F_{,z}\left( z\right) $ grows with growing $\mu .$ The vector field $\phi \left( z\right) $, the rate of evolution $F_{,z}\left( z\right) ,$ and the metric function $F\left( z\right) $, found numerically for $\phi_{,z} \left( 0\right)=0,$  $d_{0}=3,$ $\mu =10,$ $F_{0}=-4,$ and $\Omega =0.001$, are shown in the Figure 8. For this set of parameters, in accordance with (\ref{F(x^0*)=0}),\ the today's ``date'' is $z^{\ast }=Hx^{\ast }=3.45.$ 

The damping oscillations of the vector field $\phi \left( z\right) $ give rise to the oscillations of $F_{,z}\left( z\right) $ and result in slight variations of $F\left( z\right) $ in the vicinity of the compression-to-expansion transition. Practically, for $\mu \gg 1$\ the oscillations vanish at $z>1.$
\begin{figure}
  \includegraphics[width=16 cm]{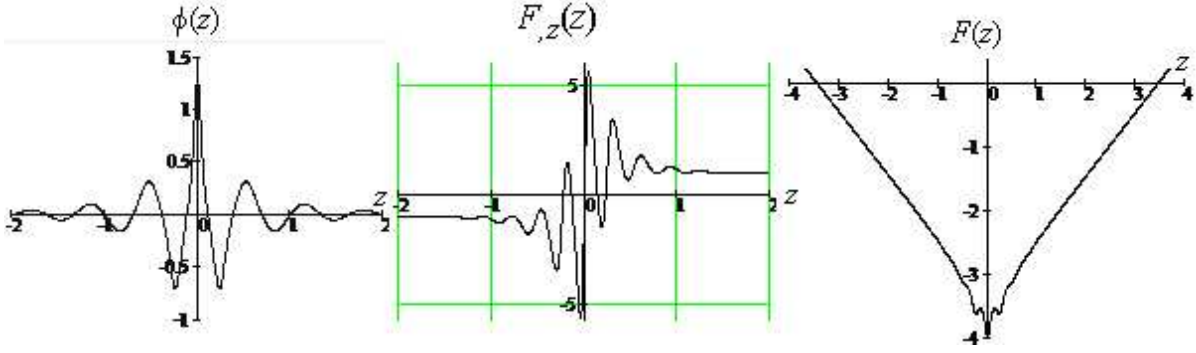}\\
  \caption{The vector field $\phi \left( z\right) $, the rate of evolution $F_{,z}\left( z\right) ,$ and the metric function $F\left( z\right) $, found numerically for $d_{0}=3,$ $\phi_{,z} \left( 0\right)=0,$ $\mu =10,$ $F_{0}=-4,$ and $\Omega =0.001$}\label{Fig8}
\end{figure}	
	
Another possible origin of the deceleration-to-acceleration switching at about the half of the age of the Universe (within our approach) can be a slight antisymmetric contribution of $\phi _{,z}\left( 0\right) \neq 0$ in the boundary conditions (\ref{bound cond with ord matter}). Numerical analysis shows, that a small nonzero $\phi _{,z}\left( 0\right) ,$ $\left| \phi _{,z}\left( 0\right) \right| \ll \left| \phi \left( 0\right) \right| ,$ can also lead to the appearance of a minimum.
	
\subsection{\label{sec44}Regular solutions with positive $\widetilde{\Lambda }$}
	
There is an important difference between the boundary conditions (\ref{boundary relations}) and (\ref{bound cond with ord matter}). The relation (\ref{boundary relations}) can be satisfied only if $\widetilde{\Lambda }<0,$ provided that $a<0.$ Appearance of the term $\Omega e^{-d_{0}F_{0}}$ in (\ref{bound cond with ord matter}) admits the solutions with positive $\widetilde{\Lambda }.$ If $\widetilde{\Lambda }$ changes sign, then $H$ (\ref{Hubble constant}) becomes imaginary. The equations (\ref{Eq for fi},\ref{F_,z,z = ...}) are invariant against $H\rightarrow iH,$ but the boundary conditions (\ref{bound cond with ord matter}) are not:
\begin{equation}\label{bound cond with ord matter Lambda positive}
    \phi _{,z}^{2}\left( 0\right) +\mu ^{2}\phi ^{2}\left( 0\right) =-1+\Omega e^{-d_{0}F_{0}},\quad F_{,z}\left( 0\right) =0,\quad F\left( 0\right) =F_{0},\quad \widetilde{\Lambda }>0.
\end{equation}
The necessary condition for regular solutions with $\widetilde{\Lambda }>0$ is the existence of an extremum moment $\left( F_{,z}\left( 0\right) =0\right) $ with the energy density of the ordinary matter exceeding the kinetic energy of expansion:
\begin{equation*}
    \Omega e^{-d_{0}F_{0}}=\frac{\varkappa \rho \left( 0\right) }{\widetilde{\Lambda }}>1,\quad F_{,z}\left( 0\right) =0,\quad \widetilde{\Lambda }>0.
\end{equation*}
In the case $\widetilde{\Lambda }>0,$ $\mu \ll 1$ the analytical solution of the equations (\ref{Eq for fi},\ref{F_,z,z = ...}) with the boundary conditions (\ref{bound cond with ord matter Lambda positive}) is
\begin{equation*}
    F\left( z\right) =F_{0}+\frac{1}{d_{0}}\ln \left[ \left( 1-\frac{1}{2}\Omega e^{-d_{0}F_{0}}\right) \cos \left( d_{0}z\right) +\frac{1}{2}\Omega e^{-d_{0}F_{0}}\right] ,
\end{equation*}
\begin{equation*}
    \phi \left( z\right) =\mu ^{-1}\left[ \left( 1-\frac{1}{2}\Omega e^{-d_{0}F_{0}}\right) \cos \left( d_{0}z\right) +\frac{1}{2}\Omega e^{-d_{0}F_{0}}\right] ^{-1}.
\end{equation*}
The scale factor $R\left( z\right) $ and the rate of evolution $F_{,z}\left( z\right) ,$
\begin{equation*}
    R\left( z\right) =e^{F_{0}}\left[ \frac{1}{2}\Omega e^{-d_{0}F_{0}}-\left( \frac{1}{2}\Omega e^{-d_{0}F_{0}}-1\right) \cos \left( d_{0}z\right) \right] ^{\frac{1}{d_{0}}},
\end{equation*}
\begin{equation*}
    F_{,z}\left( z\right) =\frac{\sin \left( d_{0}z\right) }{\left( 1-\frac{2}{\Omega }e^{d_{0}F_{0}}\right) ^{-1}-\cos \left( d_{0}z\right) },
\end{equation*}
are periodical functions with no singularities, see red curves in Figure 9. For the values of the parameters $d_{0}=3,$ $\mu =0.02,$ $F_{0}=-3,$ $\Omega e^{-d_{0}F_{0}}=1.032$ there is no difference between the curves found numerically and analytically.
\begin{figure}
  \includegraphics[width=16 cm]{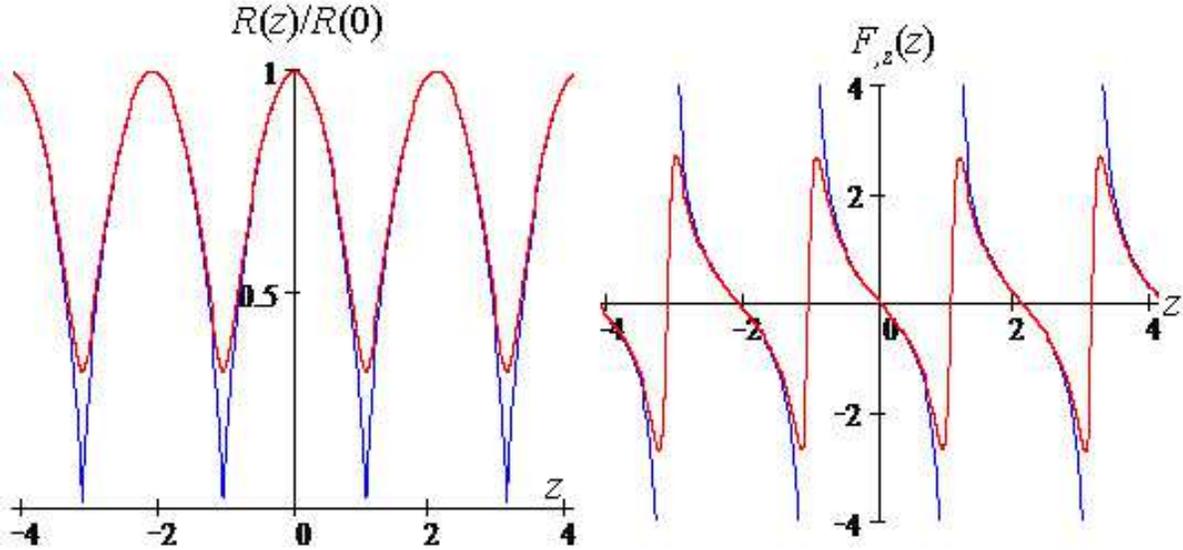}\\
  \caption{Left: scale factor $R(z)$/$R(0)$. Right: rate of evolution $F_{,z}(z)$. Red curves -- numerical (coinciding with analytical) solutions for $d_{0}=3,$ $\mu =0.02,$ $F_{0}=-3,$ $\Omega e^{-d_{0}F_{0}}=1.032$; blue curves -- (\ref{R(z)  no field}) and
   (\ref{F_,z (z)  no field}),  $\Omega e^{-d_{0}F_{0}}=1$.} \label{Fig9}
\end{figure}	

Without the massive field $\left( \phi =0\right) $ the solutions with positive $\widetilde{\Lambda }$ \ are possible only if the parameters are fine tuned $\left( \ \varkappa \rho \left( 0\right) =\widetilde{\Lambda }\right) :$
	
\begin{equation*}
    F\left( z\right) =F_{0}+\frac{1}{d_{0}}\ln \cos ^{2}\frac{d_{0}z}{2},
\end{equation*}
\begin{equation}\label{F_,z (z)  no field}
    F_{,z}\left( z\right) =-\tan \frac{d_{0}z}{2},
\end{equation}
\begin{equation}\label{R(z)  no field}
    R\left( z\right) =e^{F_{0}}\cos ^{\frac{2}{d_{0}}}\frac{d_{0}z}{2}.
\end{equation}

The scale factor $R\left( z\right) $ (\ref{R(z) no field}) and the rate of expansion $F_{,z}\left( z\right) $ (\ref{F_,z (z) no field}) are presented in Figure 9 (blue curves). These ''fine tuned'' $\left( e^{F_{0}}=\left( \frac{2\varkappa \rho _{0}}{d_{0}\left( d_{0}-1\right) H^{2}}\right) ^{1/d_{0}}\right) $ singular solutions have periodical singularities at $z=z_{n}=\frac{\pi }{d_{0}}\left( 1+n\right) ,$ $n=\pm 1,\pm 2,...$ . In the vicinity of each singular point $z_{n}=Hx_{n}^{0},$ as well as at $H\rightarrow 0,$ the Hubble constant $H$ drops out, and the scale factor (\ref{R(z) no field}) (in the ordinary units $x^{0}=z/H$) reduces to
\begin{equation}\label{R(x^0) =   FRW}
    R\left( x^{0}\right) =\left( \frac{d_{0}\varkappa \rho _{0}}{2\left( d_{0}-1\right) }\right) ^{1/d_{0}}\left| x^{0}-x_{n}^{0}\right| ^{2/d_{0}},\quad \left| \frac{x^{0}}{x_{n}^{0}}-1\right| \ll 1.
\end{equation}
At $d_{0}=3$ (\ref{R(x^0) = FRW}) reproduces the scale factor of the Friedman-Robertson-Walker \cite{FRW cosmology} cosmology with dust matter in the plane space geometry. The longitudinal vector field $\ \phi \neq 0$ removes the singularities, see red curves in the Figure 9.
	
The idea of ``oscillating Universe'' is actively supported by Lessner\cite{Lessner} as an alternative approach to cosmology. Meanwhile, I would rather not call it alternative, for all the above solutions, including the oscillating ones, are derived completely within the frames of the standard Lagrange approach and General relativity.

\section{\label{section5}Summary}
	
 Big Bang singularity is not the inevitable property of the Universe evolution. It is a consequence of our not knowing the physical nature of the dark sector, including the origin of its ability to resist compression. Nevertheless, there is a possibility for macroscopic description of the Universe evolution within the frames of the Einstein's theory of general relativity. The simplest non-gauge vector field with the Lagrangian
 \begin{equation}\label{Lag case a}
    L=a\left( \left( \phi _{;M}^{M}\right) ^{2}-m^{2}\phi ^{K}\phi _{K}\right) -V_{0}
 \end{equation}
can be that missing link in the chain, necessary to understand the mechanism of accelerated expansion of the Universe, and avoid, better say -- resolve, the Big Bang singularity.
	
It is rather involuntarily, but the modern interpretations of the observational data are based on the idea of the Big Bang birth of the Universe. The cosmic background radiation, among other phenomena, definitely testifies that the Universe had been strongly compressed in the past. But how strongly? That is the question. The information from the past, coming to us with electromagnetic waves, tells us only about the phenomena that happened after the Universe became transparent. The far extrapolation to the Plank's era is based on the assumption that the singularity is the inevitable property of cosmological solutions of the Einstein equations. It is so for the solutions, taking into account only the ordinary matter (including electromagnetic radiation). The discovery of the accelerated expansion strictly pointed on the existence of hidden sector, able to resist the compression. Though the presented above solutions are regular, they do not restrict the value of compression from above. The degree of maximum compression is determined by a free parameter $F_{0}.$ At large negative $F_{0}$ the regular transition from compression to expansion looks like inflation. However, one should keep in mind that the dust matter approximation is applicable if the galaxies are located at far distances from one another.
	
The macroscopic theory can help understanding some specific features of the Universe evolution, but it does not determine the physical nature of what we call ``dark sector''. The vector fields can be considered as appropriate tools for macroscopic description of the Universe evolution.\ From my point of view, it is more important what the fields do, than how we call them. The previous analysis confirms that the zero-mass vector field is adequate for the description of dark energy\cite{Meier3}. In some papers the dark energy is considered as the origin of acceleration. However, zero-mass fields can be responsible for contractions and expansions only at a constant rate. The acceleration is connected with massive vector fields. Being massive, they can not be associated with dark energy. At the same time, their energy-momentum tensor differs from the one of the ordinary matter. Therefore, the massive fields should correspond to dark matter (at least, partly).
	
From the General relativity viewpoint, all three kinetic terms in the Lagrangian (\ref{Lagrangian sym and antisym}) have equal rights. In flat spacetime the antisymmetric term $\sim \left( b-c\right) $ corresponds to electromagnetic field (photons). The symmetric term $\sim \left( b+c\right) $ has attention to the gauge vector particles\cite{Bogolubov-Shirkov}. There is no reason why the term $\sim a$ should be ``more equal than others''. So, the Lagrangian (\ref{Lag case a}) also deserves to be associated with some particles, existing in nature. In accordance with the Subsection ``Regular cosmological solutions'', the observed\cite{Suzuki} point of minimum $z_{\min },$ where the deceleration turns back to acceleration (see Figures 5,6), could correspond to $\mu \sim $ $1$. It is worth trying to detect an extremely light particle with the rest energy $mc^{2}\sim \hbar H\sim 10^{-33}$ eV. The trouble is that the massive vector field displays itself via gravitation (it curves the spacetime), but there is no evidence of its direct interaction with the ordinary matter.\


\begin{thebibliography}{9}
\bibitem{FOOTNOTE} In the derivations below the units with $c=1,$ and $\hbar =1$ are used.
\end{thebibliography}
\end{document}